\documentstyle[12pt,fleqn]{article}
\newcommand{\nwc}{\newcommand}

\newcommand{\be}{\begin{equation}}
\newcommand{\ee}{\end{equation}}
\newcommand{\nn}{\nonumber}
\newcommand{\beba}{\begin{equation}\begin{array}{lcl}}
\newcommand{\eaee}{\end{array}\end{equation}}
\newcommand{\bea}{\begin{eqnarray}}
\newcommand{\eea}{\end{eqnarray}}
\newcommand{\ba}{\begin{array}}
\newcommand{\ea}{\end{array}}

\newcommand{\ns}{\normalsize}
\newcommand{\refs}[1]{(\ref{#1})}
\newcommand{\ler}{\stackrel{\scriptstyle <}{\scriptstyle\sim}}
\newcommand{\ger}{\stackrel{\scriptstyle >}{\scriptstyle\sim}}

\nwc{\ra}{\rightarrow}
\nwc{\lra}{\longrightarrow}
\nwc{\lera}{\leftrightarrow}
\nwc{\lolera}{\longleftrightarrow}
\nwc{\pa}{\partial}
\nwc{\pri} {^{\prime}}
\nwc{\dpr} {^{\prime\prime}}
\def\a{\alpha}

\def\g{\gamma}

\def\f{\phi}
\def\fkbp{\dot{\phi}^{\bar k}\kern-1em \bar{\raisebox{-0.1em}{\phantom{X}}}}
\def\fmbp{\dot{\phi}^{\bar m}\kern-1.25em
  \bar{\raisebox{-0.08em}{\phantom{X}}}}

\def\n{\nu}

\def\p{\pi}

\def\r{\rho}

\def\D{\Delta}
\def\F{\Phi}
\def\G{\Gamma}

\def\cl{{\cal L}}

\def\zz{\relax{\sf Z\kern-.3em Z}}
\def\ZZ{\relax{\sf Z\kern-.4em Z}}
\def\ZZZ{Z\kern -0.28em Z}
\def\CC{{\rm \kern .25em
             \vrule height1.4ex depth-.12ex width.06em\kern-.31em C}}

\def\Sb{\bar{S}}
\def\Tb{\bar{T}}
\input epsf

\begin{document}
\begin{titlepage}
\title{{\large\bf Post Inflationary Behaviour of String Moduli}\\
                          \vspace{-4cm}
                          \hfill{\ns TUM-HEP 244/96\\}
                          \hfill{\ns SFB-375/96\\}
                          \hfill{\ns hep-th/9604096\\[.1cm]}
                          \hfill{\ns \today}
                        \vspace{2cm} }

\author{Andr\'e Lukas
        \thanks{Email: alukas@physik.tu-muenchen.de}~
        \thanks{Address after March 1996: Department of
        Physics, University of Pennsylvania,
        Philadelphia, PA 19104, USA.}~,
        Alexander Niemeyer~\thanks{Email: niemeyer@physik.tu-muenchen.de}\\
        {\ns and}\\[2mm]
        Masahiro Yamaguchi~\thanks{Email: myamaguc@physik.tu-muenchen.de}~
        \thanks{On leave of absence from Department of Physics, Tohoku
        University, Sendai 980-77, Japan}
        \\[1cm]
        {\ns Physik Department}\\
        {\ns Technische Universit\"at M\"unchen}\\
        {\ns D-85747 Garching, Germany}\\}

\date{}
\maketitle

\begin{abstract} \baselineskip=6mm
We analyze the behaviour of moduli fields in string effective models between
the end of inflation and reheating. The effective moduli potential during this
era is derived for a class of simple models. We argue that this potential
significantly stabilizes the modulus at its high energy minimum, if some
restrictions on modular weights are met. Two mechanisms to further stabilize
the moduli to their low energy minima are discussed explicitly: coinciding
minima at a point of enhanced symmetry, and the smooth transition from
high to low energy minimum by an effective mass term $C^2 H^2$. For both
cases we present explicit examples, and $C^2$ is found to be $O(10)$ at
most. In addition, we show that during a smooth transition the reduction of the
modulus amplitude strongly depends on the shape of the low energy potential.
\end{abstract}

\thispagestyle{empty}
\end{titlepage}

%%%%%%%%%%%%%%%%%%%%%%%%%%%%%%%%%%%%%%%%%%%%%%%%%%%%%%%%%%%%%%%%%
%%% Introduction %%%%%%%%%%%%%%%%%%%%%%%%%%%%%%%%%%%%%%%%%%%%%%%%
%%%%%%%%%%%%%%%%%%%%%%%%%%%%%%%%%%%%%%%%%%%%%%%%%%%%%%%%%%%%%%%%%

\section{Introduction}

In view of the current understanding of superstring theory, moduli fields are
likely to have masses of the gravitino mass scale, which is assumed to be at,
or at least not far from, the electroweak scale.  The existence of such very
weakly interacting scalar fields with an electroweak--scale mass would cause a
cosmologically disastrous problem, which is referred to as the (cosmological)
moduli problem in the context of string theory~\cite{mod_prob}. This
problem can be regarded as a string version of the notorious Polonyi
problem~\cite{pol_prob} associated with a hidden sector supersymmetry
breaking in supergravity. The most severe consequence is that late moduli
decay after coherent oscillation would totally change the abundances of light
nuclei successfully predicted by the standard big--bang nucleosynthesis
theory.

\vspace{0.4cm} 

A number of ideas to solve the moduli problem (and the Polonyi problem) have
been proposed over the last decade, some of which have been reviewed in
ref.~\cite{mod_rev}. What makes the problem particularly serious is the fact
that unlike most other unwanted relics the energy stored in moduli cannot
be diluted away by ordinary inflation. At the end of inflation moduli will
generically be displaced by $O(1)$ in Planck units from their low energy
minimum and finally (once the Hubble constant approaches their mass)
start late decaying coherent oscillations with the aforementioned disastrous
consequences.  Two types of (cosmological) solutions have been proposed in view
of this particular problem. The first assumes a second late period of
inflation~\cite{weak_inf} (e.~g.~at the weak scale) of a few e--foldings \
sufficiently long to dilute away the energy density stored in the moduli.
A very interesting realization of this mechanism is provided by the idea
of thermal inflation~\cite{th_inf}.

In this paper we will be concerned with the second type of solutions which
attempts to stabilize the moduli to their minimum before the coherent
oscillations start. A useful observation towards such a solution is that
all fields receive an effective mass of the order of the Hubble parameter $H$
due to the energy density present in the universe~\cite{eff_pot}. This mass,
which will dominate in early stages can stabilize moduli very effectively to a
minimum. Unfortunately, this minimum -- the high energy
minimum -- will generally not coincide with the low energy minimum and
oscillations reappear as soon as the moduli start to roll towards the second.
It has, however, been suggested that symmetries might provide a reason for
the low and high energy minima to coincide~\cite{symm}.

In string theory the role of this symmetry can be played by modular
invariance with the minimum sitting at a point of enhanced symmetry.
In fact, a self--dual point is an extremum of the potential and the moduli
problem can be solved once this extremum is arranged to be a minimum in
each stage of the cosmological evolution. Though possible in principle,
this scenario poses a problem when it is realized during the inflationary
epoch: In models based on low energy effective string actions inflation
is generally difficult to obtain~\cite{st_inf} and it appears to be problematic
to construct explicit examples as long as the precise mechanism
of inflation is unknown. Not a single consistent
example with an inflationary behaviour which stabilizes a modulus to a point
of enhanced symmetry is known to us.

Recently Linde~\cite{linde} has  proposed a dynamical damping mechanism to
reduce the amplitude of the moduli which does not assume coinciding high and 
low energy minima. The crucial observation is that an effective moduli
mass term $m_{T, {\rm eff}}\simeq C^2H^2/2$ with a large number $C^2\gg 1$
causes a significant damping of the oscillation amplitude. This happens
because such a large value $C^2$ guarantees a `smooth' transition of
the modulus from the high to the low energy minimum at the time when $H$
has decreased down to the low energy modulus mass $m_T$ and the low energy
potential becomes relevant. In ref.~\cite{linde} it has been shown that this
damping mechanism works for a quadratic low energy potential
$V_0\simeq m_T^2T^2/2$ and leads to a damping of the oscillation amplitude
by a factor of $\sim\exp (-\p pC/2)$ where $p$ described the expansion rate
of the universe via $H=p/t$.

\vspace{0.4cm}

In this paper, we would like to study the moduli potential in the post
inflationary era with particular emphasis on the aforementioned possibilities
to solve the moduli problem. In particular, we will derive the effective
scalar potential for a modulus which describes its behaviour during the
preheating phase after inflation, namely the era when the inflaton obeys a
damped oscillation until it decays to reheat the universe. It is this era
that determines the initial condition of the moduli oscillation.
Therefore, for a solution of the moduli problem
based on a symmetry induced minimum this era has to be considered necessarily.
Moreover, we do not have to deal with consistency problems which arise
from the unknown mechanism of inflation in string effective models if the
stabilization of moduli is implemented in the preheating phase. All we
have to assume is the existence of such a phase with the energy density
being dominated by a coherently oscillating field which -- depending on
the type of inflation -- might or might not coincide with the inflaton.
Indeed, we will construct an explicit example which shows the stabilization
of a modulus at a self--dual point during this era.

Also the mechanism of ref.~\cite{linde} mentioned above can be addressed
in our framework. A solution of the gravitino problem~\cite{gravitino} implies
a low reheating 
temperature $\ler 10^9 $ GeV, corresponding to the Hubble parameter $\ler 1$ 
GeV, well below the electroweak scale. Since the smooth transition between
high and low energy minima occurs once $H$ drops to $H\sim m_T$ it will
typically take place during the preheating era. In our framework, we are also
able to explicitly compute the value $C^2$ which was put in by hand in
ref.~\cite{linde} and it can be checked whether it can be really made as
large as required.

\vspace{0.4cm}

To illustrate the above ideas, we will concentrate ourselves on a simple
class of models consisting of one modulus $T$ with a low energy potential
$V_0$ and an oscillating field (inflaton) $\F$. They will be introduced in
the next section where a general analysis of their properties during
preheating is presented. In particular, we derive an effective modulus
potential by performing a short time average over one period of the $\F$
oscillation. In section 3 we will focus on string motivated
models: The low energy potential is assumed to be dual and to originate from
gaugino condensation~\cite{gau_con_gen} and the (modulus dependent) Yukawa
coupling of $\F$ is chosen in accordance with modular invariance. For these
models we will first discuss the mechanism of stabilization at the high
energy minimum in some detail. It turns out that this first stabilization
(which is essential for the other mechanisms to work) is possible for
a restricted range of modular weights of $\F$ only. Then we construct an
explicit example with coinciding high and low energy minima at the self--dual
point. As for Linde's mechanism, we first perform a `phenomenological'
analysis for $T$--dual low energy potentials. We find that the conclusions
concerning a suppression of moduli oscillations can differ substantially
from the case of a purely quadratic potential considered in ref.~\cite{linde}.
After this, information about the possible values $C^2$ is extracted.
It turns out that large values are {\em not} associated with {\em large} but
with {\em steep} couplings in the theory. In this respect the properties
of modular functions naturally present in string effective models turn out
to be very interesting. The results will be summarized and commented on in
section 4.

\section{The general framework}

In this section we would like to present a class of simple models suitable
to describe the behaviour of moduli fields during the preheating period.
We will investigate their general properties and in particular derive an
effective moduli potential which incorporates the effect of the energy
density stored in the coherent oscillations.
Explicit examples will be given in the next section.

Our models consist of two chiral superfields, namely a modulus $T$ and the
oscillating field $\F$. We do not restrict ourselves to any specific type
of inflation which precedes the preheating phase. For chaotic
inflation~\cite{chaotic} $\F$ typically will be the inflaton itself
whereas for hybrid inflation~\cite{hybrid} a false vacuum expansion is
followed by oscillations of $\F$.

The model is specified by a K\"{a}hler potential and a superpotential of the 
following form\footnote{Chiral
  superfields and their scalar components will be denoted by the same
  symbol.}: 
\be \ba{lll}
K &=& K_0(T,\bar{T})+Z(T,\bar{T})|\F |^2 \\
W &=& W_0(T)+\frac{1}{q}F(T)\F^q \ea \ .
 \label{model}
\ee 
Here the K\"{a}hler potential $K_0$ of the modulus and the metric $Z$ of $\F$
are kept general. In the examples we will concentrate on string 
motivated expressions for these quantities. Throughout this paper, we use
Planck units $M\simeq 2.4\times 10^{18} \mbox{GeV} =1$.  We have
neglected higher $\Phi$ terms in eq.~(\ref{model}) which may dominate the
potential during inflation, because we focus on the preheating phase where
$\Phi <1$.  For simplicity, we assume that the $\Phi$ field oscillates
around the origin. The superpotential has been split into two parts: a modulus
potential $W_0$ originating from nonperturbative effects like gaugino
condensation and a $\F$--term with a modulus dependent coupling $F$. We
allow the power $q$ to take the values $q=2,3$ corresponding to a mass term or
to a Yukawa coupling for $\F$.  Let us denote the intrinsic mass scales in
$W_0$ by $m_T = O(W_0)$ and in $F$ by $M_\F = O(F)$. To be consistent with the
constraint on the anisotropy of the photon background the mass
(coupling) $M_\F$ should be well below the Planck scale, typically $M_\F\simeq
10^{-5}$--$10^{-6}$ for chaotic inflation.  For hybrid
inflation, $M_{\Phi}$ can be taken somewhat smaller.  We will assume throughout
this discussion that the low energy modulus mass $m_T$ is much smaller,
i.~e.~$m_T\ll M_\F$. For moduli potentials based on gaugino condensation we
have $m_T\sim m_{3/2}\sim 10^{-15}$ so that this can be safely assumed.

\vspace{0.4cm}

Generally, the behaviour of the model~\refs{model} is expected
to be complicated. It has to be analyzed in the full field space of four
real fields and using the complete scalar potential. Having the ordinary
scenario of reheating in mind one might think about the following situation:
A rapidly oscillating $\F$ dominates the energy density of the universe
and the modulus is moving in this given `background'. Then an effective
potential for the modulus can  be obtained by a short time average
over one period of the $\F$ oscillation. Such an average procedure would
simplify the situation considerably. Let us now analyze under which
assumptions this picture is really correct.

\vspace{0.4cm}

As soon as $\F\ll 1$ the scalar potential can be expanded as
\bea
 V &\simeq& V_0+{\tilde V}_1 \nn \\
 {\tilde V}_1 &=& f(T,\Tb)|\F |^{2q-2} \label{potential}\\
 f &=& Z^{-1}e^{K_0}|F|^2 \ , \nn
\eea
where $V_0$ is the low energy modulus potential originating from
$W_0$. Higher powers in $\F$ have been neglected.
First we should require that the energy density is dominated by
$\F$, i.~e.~$V_0\ll {\tilde V}_1$. Then $H^2\sim {\tilde V}_1$ and the condition is
consistently fulfilled as long as
\be
 H^2\gg V_0\sim m_T^2\ . \label{cons_cond1}
\ee
Since the low energy mass $m_T$ is much smaller than the inflationary scale
$M_\F$ this condition will be fulfilled for a wide range in $H$. Even for
$H\ler m_T$ it may hold if $T$ is close enough to its low energy minimum. 
We assume that this is the case in view of the fact that we are concerned 
with a solution of the cosmological moduli problem where the amplitude of 
the moduli oscillation is either small or damped quickly. 
Self--consistency of this assumption should be checked in each
case.  From 
the expressions~\refs{potential} we can estimate the effective masses
for $T$ and $\F$ as follows:
\bea
 m_{T,{\rm eff}}^2 &=& |\frac{f_T}{f}|^2 H^2 \\
 m_{\F ,{\rm eff}}^2 &=& \frac{H^2}{|\F^2 |}\ ,
\eea
with
\be
 \frac{f_T}{f} = -\frac{Z_T}{Z} + K_{0T} +\frac{F_T}{F}\ .
 \label{fprime}
\ee
A short time average is justified if $m_{T,{\rm eff}}\ll m_{\F ,{\rm eff}}$
which in turn implies
\be
 |\F| \ll |\frac{f}{f_T}|\ . \label{cons_cond2}
\ee
Inspection of eq.~\refs{fprime} shows that this condition is usually fulfilled
once $|\F|$ is sufficiently small. However there is one important exception:
The quantity $f_T /f$ can become very large close to zeros $F(T)=0$ of the
Yukawa coupling. For the following derivation we therefore assume that
the system -- at least for a certain period after $|\F|$ became small does not
evolve towards such a point or, equivalently, that the coupling strength
$|F|$ is bounded from below in the relevant part of field space. Later on,
we will comment on what happens if this condition is violated.

Oscillation of $\F$ will dominate the energy density of the universe as long as
$H=p/t\sim \G_\F\sim T_{\rm RH}^2$ where $H=\dot{a}/a$ is the Hubble parameter,
$\G_\F$ is the decay constant of $\F$ and $T_{\rm RH}$ is the
reheating temperature. For a radiation like behaviour of $H$ we have $p=1/2$,
whereas a matter like behaviour is characterized by $p=2/3$. In view of the
gravitino problem $T_{\rm RH}\ler 10^9$ GeV.
This implies that the universe can be dominated
by $\F$ oscillations down to values of 
$H\ler 1$ GeV. In particular the start of the coherent modulus oscillations
in the low energy potential responsible for the moduli problem at $H\sim m_T$
can fall into this era.

\vspace{0.4cm}

Let us now carry out the averaging procedure to obtain an effective potential
for the modulus in the background of the oscillating field $\F$. As discussed,
we assume that the conditions~\refs{cons_cond1} and~\refs{cons_cond2} hold.
The system is described by a Lagrange function
\be
 \cl = a^3\left[ -\left(\frac{\dot{a}}{a}\right)^2+K_{i\bar{k}}\dot{\f}^i
       \fkbp - V\right]
\ee
with the K\"{a}hler metric $K_{i\bar{k}}=\partial_i\partial_{\bar{k}}K$ and the
fields denoted by $\f^i$. For simplicity, we have assumed a flat
Robertson--Walker universe. The conjugate momenta are given by
\be
\p_i = a^3K_{i\bar{k}}\fkbp ,
\ee
and the equations of motion can be written in the form
\be
 \frac{d}{dt}\left( K_{i\bar{k}}\dot{\f}^i\right)+3HK_{i\bar{k}}\dot{\f}^i
 -(\partial_{\bar{k}}K_{i\bar{m}})\dot{\f}^i\fmbp
  +\partial_{\bar{k}}V = 0 \ . \label{eom1}
\ee
Sometimes it is useful to normalize the second derivative term in
eq.~\refs{eom1} to get
\be
 \ddot{\f}^i+3H\dot{\f}^i+\G_{jk}^i\dot{\f}^i\dot{\f}^j+\partial^i V = 0\ ,
 \label{eom2}
\ee
with the K\"{a}hler connection $\G$ defined by
$\G_{ij}^k = K^{k\bar{l}}\partial_i K_{j\bar{l}}$. Furthermore, the
Friedmann equation reads
\be
 H^2 = K_{i\bar{k}}\dot{\f}^i\fkbp+V\ . \label{constraint}
\ee
A straightforward computation under the above assumptions and
using eq.~\refs{eom1} leads to
\be
 \frac{d}{dt}\left(\p_\F\F\right)\simeq a^3(U-(q-1){\tilde V}_1)\ ,
 \label{average}
\ee
where $U = K_{i\bar{k}}\dot{\f}^i\fkbp$ is the
kinetic energy. The short time average (which we denote by $<..>$ in the
following) of  $d(\p_\F \F )/dt$ approximately vanishes and we have
\be
 <U>\simeq (q-1)<{\tilde V}_1>\ .
\ee
With the energy density and the pressure
$\r = <U +V>$, $P = <U-V>$ we derive an equation of state
\be
 P = \n\r\ ,\quad\quad \n=\frac{q-2}{q}\ .
\ee
and the Hubble parameter behaves like
\be
 H=\frac{p}{t}\ ,\quad\quad p=\frac{2}{3(1+\n )}=\frac{q}{3q-3}\ .
\ee
As expected, for $q=2$ (quadratic $\F$ potential) the expansion corresponds to
a matter dominated universe and for $q=3$ (quartic $\F$ potential) it
corresponds to a radiation dominated universe. Since $<{\tilde V}_1>=H^2/q$ we
can carry out the following replacement in the EOM for $T$
\be
 <\frac{\partial {\tilde V}_1}{\partial \Tb}> =
   \frac{1}{q}\frac{f_{\Tb}}{f} H^2\; .
\ee
This leads to the effective EOM for $T$ during preheating
\be
 \ddot{T}+3H\dot{T}+\frac{K_{0TT\Tb}}{K_{0T\Tb}}\dot{T}^2+ 
  K_{0T\Tb}^{-1}\frac{\partial V_{\rm eff}}{\partial \Tb} 
= 0\ ,
  \label{eff_eom}
\ee
with the potential $V_{\rm eff}$ given by
\be
 \ba{lll}
  V_{\rm eff} &=& V_0+V_1= V_0+\frac{1}{q}\; g\; H^2 \\ [0.3cm]
  g &=& \ln f = -\ln Z + K_0 + \ln |F|^2 \ .
 \ea \label{eff_pot}
\ee

For $H\gg m_T$ the potential~\refs{eff_pot} is dominated by the second 
term originating from the $\F$ oscillations. The modulus will therefore
settle down to one of the minima $T_1$ of $g$.
This value can be assumed as an initial value for the further development of
$T$ which sets in once the low energy potential becomes important, i.~e.~once
$H$ approaches $m_T$. In the spirit of ref.~\cite{linde} we can introduce
a parameter $C^2$
\bea
  C^2(T) &=& \frac{1}{q}\frac{\normalsize g_{\Tb}}{\normalsize T-T_1} \nn \\
  g_T &=& -\frac{\normalsize Z_T}{\normalsize Z}+K_{0T}+\frac{\normalsize F_T}{
  \normalsize F}\ . \label{C_def}
\eea
Then the EOM can be rewritten as
\be
 \ddot{T}+3H\dot{T}+\frac{K_{0TT\Tb}}{K_{0T\Tb}}\dot{T}^2+K_{0T\Tb}^{-1}
 \left(\frac{\partial V_0}{\partial \Tb}+C^2(T)H^2
 (T-T_1)\right) = 0\ . \label{eff_eom_C}
\ee
Notice that, in contrast to ref.~\cite{linde}, $C^2$ depends
on $T$ and can -- along with the starting value $T_1$ -- be computed
explicitly once a specific model has been chosen.

\vspace{0.4cm}

Let us now discuss the general consequences of this result before we turn
to concrete models. Of course it might happen that the system runs into a
region of very low coupling $|F(T)|$ even before the above stage is
reached, i.~e.~before $|\F|$ is sufficiently small. In this case which strongly
depends on the initial conditions there is no chance to generate a
substantial damping of the modulus and we are left with the usual moduli
problem.

If however the system evolves according to the above scenario, a damping
of the modulus oscillation might result. First we would like to analyze
the stabilization of $T$ to the high energy minimum $T_1$. Clearly, 
$g_T(T_1)=0$, and for $T_1$ to be a stable minimum the requirement
\be
 g_{T\bar{T}}(T_1) > |g_{TT}(T_1)| \label{stability}
\ee
has to be fulfilled. In particular $g_{T\bar{T}}>0$, which might impose
severe restrictions on the model since $g_{T\bar{T}}$ does not depend on the
superpotential but on the K\"ahler potential only. The stability
condition~\refs{stability} can be used to derive an upper bound on the
value of $C_1^2=C^2(T_1)$ which determines how effective the stabilization
to $T_1$ can be. From eq.~\refs{C_def} we have
\be
 C^2(T\simeq T_1) \simeq \frac{1}{q}\left[ g_{T\bar{T}}(T_1)+
                  \frac{\bar{T}-\bar{T}_1}{T-T_1}g_{TT}(T_1)\right]
 \label{C1}
\ee
and with condition~\refs{stability} it follows that
\be
 |C_1^2|\le \frac{2}{q}g_{T\bar{T}}(T_1) \ . \label{C_constraint}
\ee
Moreover, the order of $|C_1^2|$ will be given by the RHS of
eq.~\refs{C_constraint}. Notice that again this value depends on the
K\"ahler potential of the theory only. The implications of
eqs.~\refs{stability} and \refs{C_constraint} for string motivated models and
the stabilization process will be studied in more detail in the next section.

For the moment we assume that $T$ has been driven to $T_1$ with a certain
precision and starts its further evolution from this point. If $T_1$
coincides with the low energy minimum $T_0$ the stabilization is
definite and moduli oscillations are suppressed.

Suppose that $T_1$ and $T_0$ do not coincide. Then, once $H\sim m_T$ the
modulus will feel the low energy potential and will start oscillations
around the true minimum $T_0$. The moduli problem simply reappears. In
ref.~\cite{linde} it has been shown that assuming a quadratic low energy
potential these oscillations can be suppressed substantially for large
values of $C^2$ by a factor of order $\exp (-\p pC/2)$. This mechanism comes
into operation as soon as $H$ drops to $H\sim m_T$. Therefore, it may play a
role for the models considered here if $T_{\rm RH}\ler m_T^{1/2}$ so that
domination of the $\F$ oscillations is maintained sufficiently long. Such
a low value of the reheating temperature is also required in order to solve
the gravitino problem.
 
Large values of $C^2$ cannot be obtained by simply introducing a large
coupling in the theory. As eq.~\refs{C_def} shows a coupling constant in
front of $F$ or $Z$ simply drops out. Instead, what one needs is a
very {\em steep} coupling such that $F_T/F$ becomes large. Since generally
$C^2$ is proportional to $V^{\prime}/V$ the validity of this statement is
not restricted to our particular class of models. In the next section we
will discuss how the conclusions of ref.~\cite{linde} change for
$T$--dual low energy potentials and to what extent large values of $C^2$ can
be obtained in string motivated models.

Finally, the modulus might run into a region with vanishing Yukawa coupling
$|F(T)|$ even after our effective potential description becomes valid. As can
be seen from eq.~\refs{eff_pot} these points show up in 
the effective potential as singularities $V_{\rm eff}\rightarrow -\infty$.
If such zeros exist the system might therefore easily be attracted to them.
To analyze the qualitative behaviour close to such a point $T_2$ let us
expand $F \simeq F_1^{\pri} (T-T_2)$. Then, according to the
condition~\refs{cons_cond2} our averaging procedure breaks down once
$1/C^2\sim |T-T_2|\sim |\F |$ and $T$ will start to oscillate around $T_2$
with an initial amplitude of $|\F|$. Since $|\F|$ can be quite small when this
happens a large suppression of the $T$ amplitude might result. Again, this
guarantees a (at least partial) solution of the moduli problem only if
$T_2$ coincides with the low energy minimum $T_0$.

For comparison we would like to briefly mention the requirements on the
amplitude of the coherent moduli oscillation posed by the nucleosynthesis
constraint. The energy density $\rho_T$ of these oscillations at the time of
the decay of $T$ divided by the photon density $n_\g$ is given by
\be
\frac{\rho_T}{n_\g} \simeq T_{\rm RH} T_{\rm in}^2 ,
\ee
$T_{\rm in}$ being the initial oscillation amplitude and assuming a matter
dominated preheating phase.

In order not to dissociate light nuclei this quantity should be less than
typically $10^{-8}$ to $10^{-12}$ GeV (for ordinary moduli masses). For a
reheating temperature of $T_{\rm RH}=10^9$ GeV this is satisfied, if $T_{\rm
in} \ler 10^{-9} $ to $10^{-11}$ in Planck units.  

In the case of a quadratic low energy potential as used in \cite{linde} one
therefore needs values of $C^2\ger 10^2\;-\; 10^3$. 

%%%%%%%%%%%%%%%%%%%%%%%%%%%%%%%%%%%%%%%%%%%%%%%%%%%%%%%%%%%%%%%%
%%% Applications and Examples %%%%%%%%%%%%%%%%%%%%%%%%%%%%%%%%%%%
%%%%%%%%%%%%%%%%%%%%%%%%%%%%%%%%%%%%%%%%%%%%%%%%%%%%%%%%%%%%%%%%%

\section{Applications and examples}

In this section we would like to discuss the various possibilities
mentioned above more explicitly. We will concentrate on string motivated
models specified by
\be
 K_0 = -w\ln (T+\bar{T})\ ,\quad\quad Z = \frac{1}{(T+\bar{T})^n}\ .
\ee
Usually, for a modulus we have $w=3$. The modular weight of $\F$ is denoted
by $n$, i.~e.~under $T\rightarrow 1/T$ duality
$\F$ transforms as $\F\rightarrow \F /T^n$. To have a duality
invariant theory the modular weight $m$ of the Yukawa coupling $F$
should be given by $m=w-qn$. Application of eq.~\refs{eff_pot} leads
to an effective potential
\be
 g = (n-w)\ln (T+\Tb) +\ln |F|^2\ ,
\ee
and the quantity $g_T$ which determines the value of $C^2$ is given by
\be
 g_T = \frac{n-w}{T+\Tb}+\frac{F_T}{F}\ .
\ee
The potential $g$ is modular invariant if and only if $\F$ has a vanishing
weight $n=0$. Since $\F$ has been `integrated' out to arrive at the above
result this property could be expected.

\vspace{0.4cm}

The explicit form of Yukawa couplings in string theory has been computed
in a number of publications~\cite{yukawa1,orbi}. Generally, it has been
found to be consistent with modular invariance. As an illustrative class we
will consider couplings of the form
\be
 F(T)\F^q = M_\F \frac{j^l(T)}{\eta^{2m}(T)}\F^q \label{yukawa}
\ee
with the Dedekind function $\eta$ and the totally modular invariant
function $j$. By symmetry requirements $F$ could contain an arbitrary function
of $j(T)$. Here we study only the subclass \refs{yukawa} with an arbitrary
power $l$ of $j$. Yukawa couplings containing $j$--functions show up for
example in Calabi--Yau compactifications~\cite{jinyukawa}. In orbifold models
the modular weight $n$ of $\F$ has been found to be in the
range~\cite{ib_luest} 
\be
 10\ge n\ge -4 \ . \label{n_bound}
\ee

Furthermore, we concentrate on a specific form of the low energy moduli
potential, which is related to the potential one expects from the process
of T--dual gaugino condensation~\cite{cond_dual}. Assuming gaugino
condensation in a model consisting of dilaton $S$ and modulus $T$ in the
usual way, i.e.
\bea
K_0&=&-\ln (S+\Sb)-3\ln (T+\Tb) \\
W_0&=&\Omega(S)/\eta(T)^6\ , \label{W_fact}
\eea
one arrives at the scalar potential
\be
V_0(S,T)=\frac 1{T_R^3 |\eta(T)^{12}|} \left (
A(S)+B(S)\left |2\frac{T_R \eta^\prime(T)}{\eta(T)} + 1\right |^2 \right ) \ .
\ee
where $T_R=T+\Tb$. The functions $A(S)$ and $B(S)$ are given in terms of
$\Omega (S)$. Usually $A(S)=-B(S)$ ($B(S)\ge 0$ always) at the minimum
(supersymmetry is not broken by the dilaton, but by the modulus) and this
will give the famous value $T \approx 1.23$. The shape of this potential is
shown in fig.~1.

\epsfbox[-80 0 500 210]{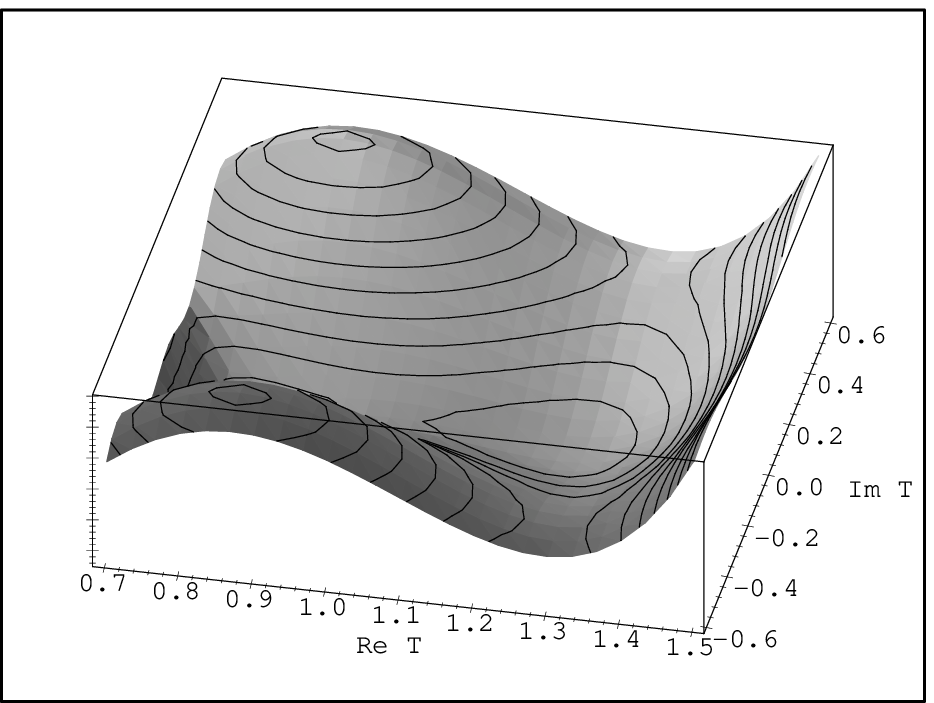}

\centerline{\em Fig 1: Gaugino condensation potential with minimum at $T=1.23$}

\vspace{0.5cm}

If $B(S)=0$, which corresponds to unbroken supersymmetry in the $T$--direction
the minimum will be at $T=\rho$ (the other self--dual point with
$\rho=\exp(i\pi/3)$). This implies that supersymmetry is unbroken by this
condensation process since for the factorizing Ansatz~\refs{W_fact}
supersymmetry cannot be broken by the dilaton. The potential looks like in
fig.~2.

\epsfbox[-80 0 500 210]{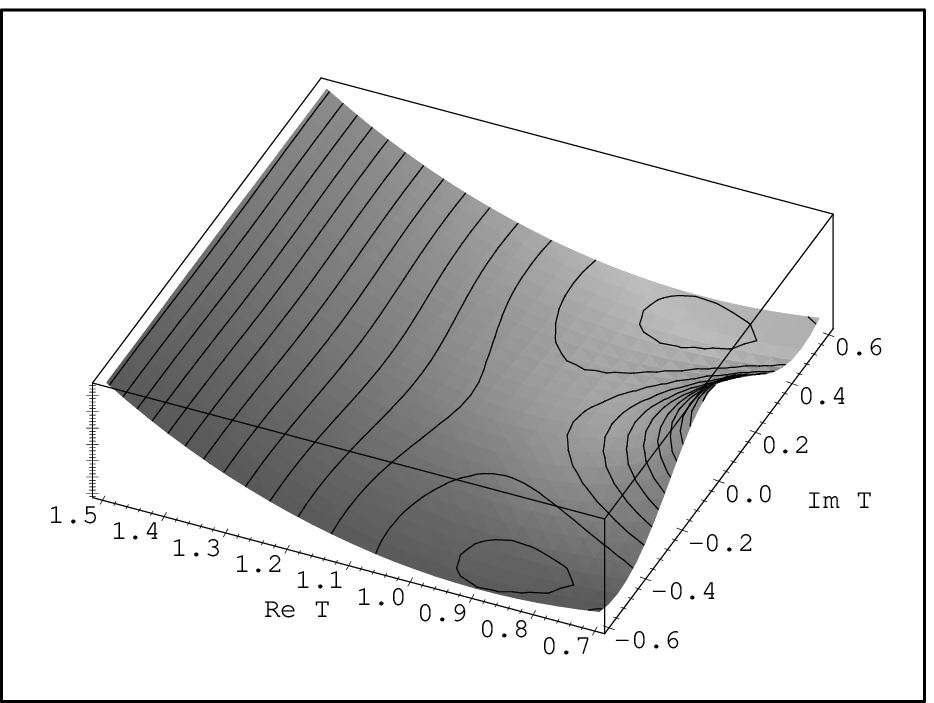}

\centerline{\em Fig 2: Gaugino condensation potential with minimum at $T=\rho$}

\vspace{0.5cm}

A stabilization of the dilaton is usually difficult to achieve~\cite{stab_dil}
and we will not address this question here. Instead, to keep our models
as simple as possible we will consider only the $T$--dependence of the
potential. Consequently, we assume that the dilaton VEV is fixed by some
unknown mechanism and $A(S), B(S)$ are thus determined.
 
Of course it is well known that the potentials discussed here generically have
a negative cosmological constant. Sometimes one can shift the vacuum energy to
zero by adjusting the value of an additional constant term in the
superpotential. We simply assume that the potentials we study
are shifted in such a way that the cosmological constant vanishes.

\subsection{Stabilization at the high energy minimum}

Now that we have specified our models let us apply the general
results of the previous section. First we study the stabilization to the high
energy minimum $T_1$, which is essential for any further suppression of moduli
oscillations. As we have seen $T_1$ is a stable minimum
of the high energy potential provided eq.~\refs{stability} holds. Since
\[
 g_{T\bar{T}} = \frac{w-n}{(T+\bar{T})^2}\ ,
\]
this implies as a necessary (but not sufficient) condition that
\be
 n < w=3\ ,
\ee
which constitutes a significant restriction on the modular weight $n$ of
$\F$. If this bound is violated, no stabilization at the high scale and
consequently no further stabilization can take
place. Furthermore, from eq.~\refs{C_constraint} we have
\be
 \left |C_1^2 \right | \le \frac{2}{q}\frac{|w-n|}{(T_1+\bar{T}_1)^2} \ .
\ee
Given the bound~\refs{n_bound} on the modular weight and the fact that the
starting value $T_1$ will be typically $O(1)$ this restricts $C_1^2$ to
a few at most. In view of this constraint it seems unlikely that the moduli
problem can be solved completely by the `$C^2$--mechanism' of
ref.~\cite{linde} though a larger value of $C^2(T)$ at $T\ne T_1$ cannot be
excluded. We will come back to this point later on.

The stabilization of $T$ to $T_1$ is governed by the EOM~\footnote{The
term proportional to $\dot{T}^2$ in eq.~\refs{eff_eom_C} can be consistently
neglected.} (cf.~eq.~\refs{eff_eom_C})
\be
 \ddot{T}+3H\dot{T}+\frac{(T_1+\bar{T}_1)^2}{w}C_1^2H^2(T-T_1)\simeq 0\; .
 \label{stab_eom}
\ee
Note that we have omitted the low energy potential $V_0$ which is unimportant
at this early stage. The solutions of eq.~\refs{stab_eom} can be written as
$(T-T_1)\sim t^{-\a}$ with
\be
 \a = \frac{3p-1}{2}\pm\sqrt{\left(\frac{1-3p}{2}\right)^2-
      \frac{p^2(T_1+\bar{T}_1)^2}{w^2}C_1^2}\ .
 \label{alpha}
\ee
Damping of the $T$ oscillations
\be
 \frac{\D T_{\rm f}}{\D T_{\rm i}} = \left(\frac{t_{\rm i}}
      {t_{\rm f}}\right)^\a\sim \left(\frac{m_T}{M_\F}\right)^\a
\ee
will be most effective if the square root in eq.~\refs{alpha} is purely
imaginary which is the case for a real and moderately large $C_1^2\sim O(1)$.
Consequently $\a$ is bounded by
\be
 \a \le \left\{\ba{lll} \frac{1}{2}&{\rm for}&p=\frac{2}{3}\\
   && \\
                      \frac{1}{4}&{\rm for}&p=\frac{1}{2} \ea\right.
\ee
and typical maximal damping ratios are $10^{-5}$ and $10^{-2.5}$ for
$p=2/3$ and $p=1/2$ respectively
(for $M_\F\sim 10^{-5}$ and $m_T\sim 10^{-15}$).
Though this suppression itself is
definitely not sufficient to solve the moduli problem for
$\D T_{\rm i}\sim O(1)$ it significantly improves a preexisting suppression
$\D T_{\rm i}\ll 1$ originating from the inflationary epoch. Of course
one has to make sure that no destabilization occurs in the following evolution
which is the problem we are going to address now.

\subsection{High and low energy minima coincide}

Models in which the minima of the high energy potential and the low energy
potential coincide, are very attractive from the viewpoint of the cosmological
moduli problem. In this case the modulus will be fixed at this minimum at early
times, when the curvature at the minimum is large as discussed in the
previous subsection. During the transition to the low energy potential it will
safely stay at that point.

In principle, this situation can be achieved by carefully tuning parameters
in the potential. Such a choice would be neither natural nor
stable under radiative corrections and is clearly unreasonable.
Therefore, coincidence of low and high energy minima should be guaranteed
by a symmetry. In a string framework duality is a very natural candidate
for this symmetry which has been proposed in ref.~\cite{symm}.

For our models, this can be realized for $n=0$ only since otherwise
duality is broken by the effective potential. This fixes the modular weight
of the Yukawa coupling $F$ to $m=w=3$. Then, by modular invariance low and
high energy potential both possess extrema at $1$ and $\r = \exp (i\p /3)$.
For a successful example, one of these points should be arranged to be a
minimum for both contributions to the potential. 

For $T_1=\rho$ 
we can make a more detailed prediction of the stabilization at the high
energy minimum. The second term in eq.~\refs{C1} vanishes at $T_1=\rho$ and the
value of $C^2$ at this high energy minimum can be explicitly computed to be
\be
 C_1^2=\frac{1}{q}\frac{w}{(T_1+\bar{T}_1)^2}=\frac{1}{q}\ .
\ee
According to eq.~\refs{alpha} this implies for the power $\a$
\be
 \a = \left\{ \ba{lll} \frac{1}{3}&{\rm for}&p=\frac{2}{3}\\
&&\\
                       \frac{1}{4}&{\rm for}&p=\frac{1}{2} \ea \right. \ ,
\ee
corresponding to a typical damping of $10^{-3.3}$ and $10^{-2.5}$ for
$p=2/3$ and $p=1/2$, respectively.

We start with an example for a minimum at $\r$. A simple Yukawa coupling
with weight $m=3$ is given by
\be
 F = \frac{M_\F}{\eta^6 (T)}\ .
\ee
The resulting effective potential $g$ has a minimum at $T_1 =\r$. As
discussed above, the low energy minimum of $V_0$ is at $\r$ if supersymmetry
is unbroken (because then $B(S)=0$).

A model where the minima coincide at the other self dual point $T=1$ can be
constructed as well. A Yukawa--coupling
\be
F= \frac {M_\F j^l(T)}{\eta^6(T)}\ , \qquad -0.068 > l > -0.016
\ee
will give the required minimum of $g$ at $T=1$. For a superpotential
of the form  $W_0=\Omega(S)/(\eta^6(T)j(T))$ instead of 
$W_0=\Omega(S)/\eta^6(T)$ the low energy minimum in $T$--direction will be 
also at $T=1$. The low energy potential of this model looks like in fig.~3.

\epsfbox[-80 0 500 210]{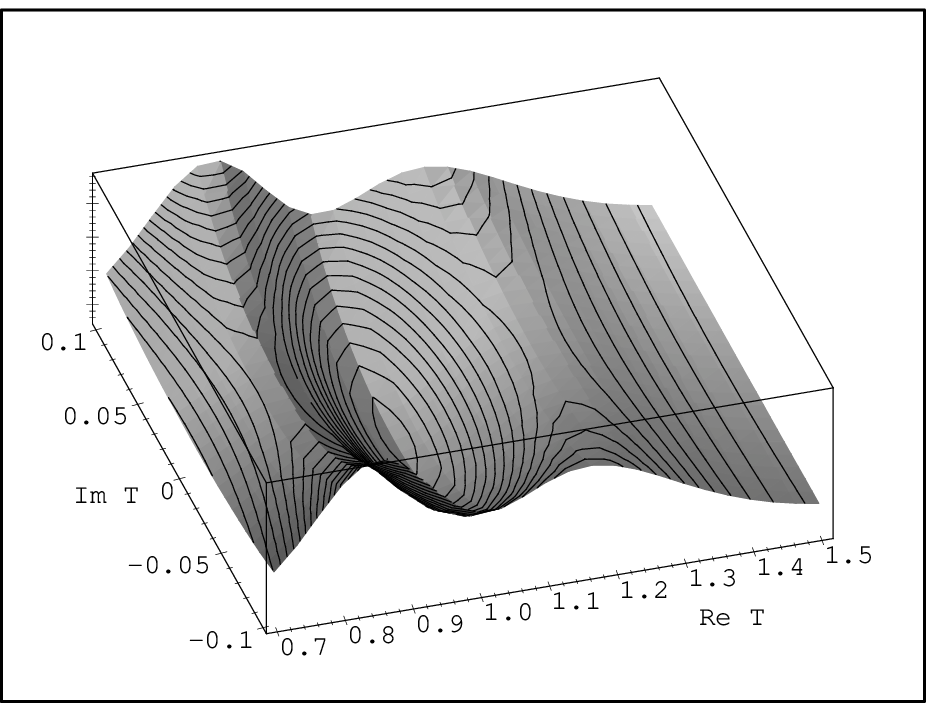}

\centerline{\em Fig 3: Potential for $W_0=\Omega(S)/(\eta^6(T)j(T))$}

\vspace{0.5cm}

Although explicit orbifold calculations do not lead to
$j$--functions~\cite{orbi}, this superpotential is perfectly consistent with
duality and might originate from Calabi--Yau compactifications, where similar
moduli dependencies show up~\cite{jinyukawa}.

Clearly, stabilization of a modulus at one of its self--dual points implies
unbroken supersymmetry in this direction.  Thus, if supersymmetry is broken
by a modulus at all at least this modulus cannot be stabilized in this way.

\subsection{High and low energy minima do not coincide}

The situation described above may not be expected for all
the (numerous) moduli of a superstring theory. Most probably not for each
modulus the minima of the low and high energy potentials will coincide. We
should therefore study this case and analyze, whether
the mechanism described in \cite{linde} will sufficiently damp the coherent
oscillations during the transition of the modulus from $T_1$ to $T_0$ (its
final minimum). First we will generalize the approach of \cite{linde} to
a more realistic low energy potential, but keep the high energy potential
quadratic. As we will see, this will already change the results
significantly.

\subsubsection{High energy minimum coincides with maximum of low energy
potential}

A smooth transition does not take place at all if the
high energy minimum coincides with a low energy maximum.

At this point we have $V^\prime_0(T_1)=V^\prime_1(T_1)=0$ and
$V_0^{\prime\prime} > 0$, $V_1^{\prime\prime} < 0$. It is clear that at all
times $V_{\rm eff}^\prime(T_1)=0$ and that at some point of time
$V_{\rm eff}^{\prime\prime}(T_1)=0$. Therefore the position of the minimum of
$V_{\rm eff}$ will not 
move smoothly from the initial to the final low energy minimum with increasing
time. Instead, the modulus will be released from its minimum at $T_1$ at the
time when $V_{\rm eff}^{\prime\prime}(T_1)=0$. Thus no reduction of the
coherent oscillation amplitude can 
result, regardless of how strong the high energy potential is 
(i.e.\ regardless of the value of $C^2$). The mechanism of
\cite{linde} cannot work in this case.

One can show that this situation occurs not only if the early minimum and the
late maximum coincide, but also when they are close to each other. The maximal
separation of the two extrema depends of course on the shape of the potentials,
but is usually rather small.

Nevertheless this behaviour gives serious restrictions for our gaugino
condensation inspired model, since the low energy potential has a maximum
(in the real direction) at the self--dual point $T=1$, if supersymmetry is broken
in the $T$-direction (as is usually the case). Therefore high energy
potentials whose minimum is at $T=1$ will not succeed in damping the coherent
oscillations of the modulus.

Fig.~4 shows the trajectory of the modulus, where $V_1=C^2 H^2 (T-1.001)^2/2$
for $C=10$ (a rather large value). It is clear that there is no damping at all,
when compared to the trajectory with $T_1=1.7$ and $C=10$, which is also given
in fig.~4. 

\subsubsection{Smooth transition between high and low energy minima}

If the minimum of $V_1$ does not lie close to a maximum of $V_0$ there will be
a smooth transition of the minimum to its final value. In \cite{linde} it was
shown that for parabolic $V_0$ and $V_1$ this can lead to a substantial damping
of the coherent modulus oscillations if $C^2$ is large enough.

In this section we study this damping in more detail for the gaugino
condensation inspired potential with the minimum at $T=1.23$ (and at the dual
point $T=0.81$). We assume the high energy potential $V_1$ to be of
parabolic form $V_1=C^2 H^2 (T-T_1)^2/2$.

\epsfbox[-80 0 500 210]{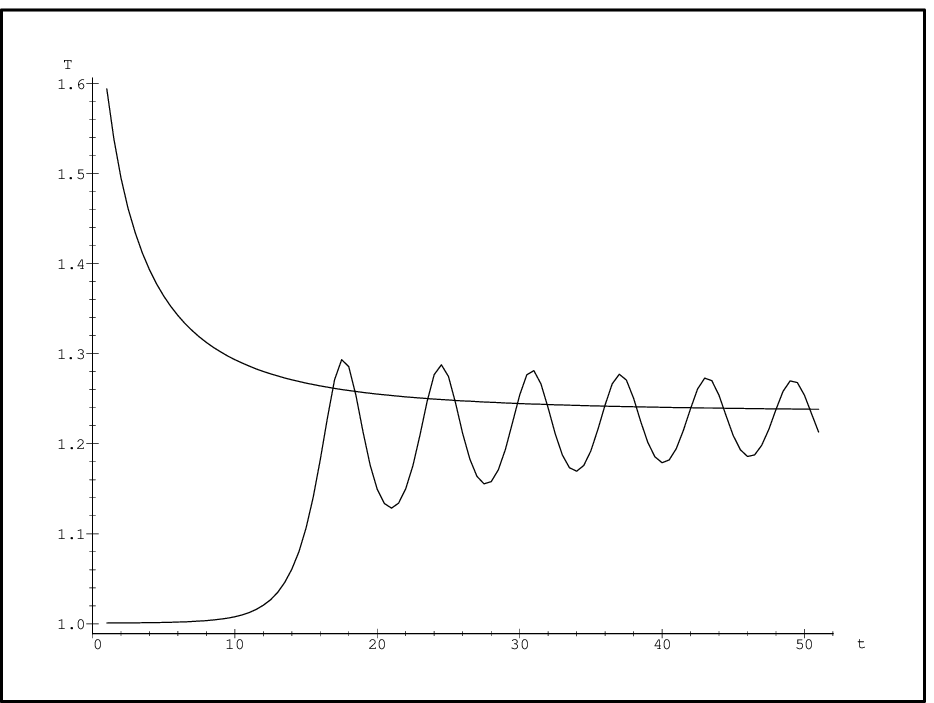}

\centerline{\em Fig 4: $T$--trajectory for $T_1=1.001$ and $T_1=1.7$ for
$C=10$} 

\vspace{0.5cm}

Unfortunately it is impossible to repeat the analytical approach of
\cite{linde} to get the  damping of the oscillation amplitude. This is, because
the more realistic potentials we are interested in do
not have the simple quadratic structure but contain the Dedekind function
$\eta$ and its derivatives. 

Trying to approximate the model by an expansion which can be solved
analytically proved to be inconsistent with the numerical solution because the
terms one has to omit become too large. Therefore we had to stick to numerical
simulations.

We have studied the dependence of the final oscillation amplitude for a set of
different starting values. The results are given in figs.~5 and 6. Each curve
represents a number of runs with the same initial minimum $T_1$ (given by the
label at the curve) and shows the resulting (normalized) amplitude for
different values of $C$ on a logarithmic scale. The curve with the label
`Linde' represents the $C$--dependence of the oscillation amplitude of the
model of \cite{linde}.

It is clear from these results that the damping of the moduli oscillations does
not only depend on the value of $C$, but also strongly on the shape of the
potential $V_0$ between the minima of $V_0$ and $V_1$. If $T_1$ is close to the
final minimum, $V_0$ can be approximated by a quadratic function. As expected,
the reduction of the amplitude with respect to $C$ is identical to the result
in \cite{linde}. However, outside of that region, there are strong
differences. If  $T_1$ falls into a region of $V_0$ which is rather flat, then
increasing the value of $C$ decreases the oscillation amplitude by a factor
which is significantly smaller. If the modulus starts at a point where $V_0$ is
steep, then the reduction factor is larger.

\epsfbox[-80 0 500 210]{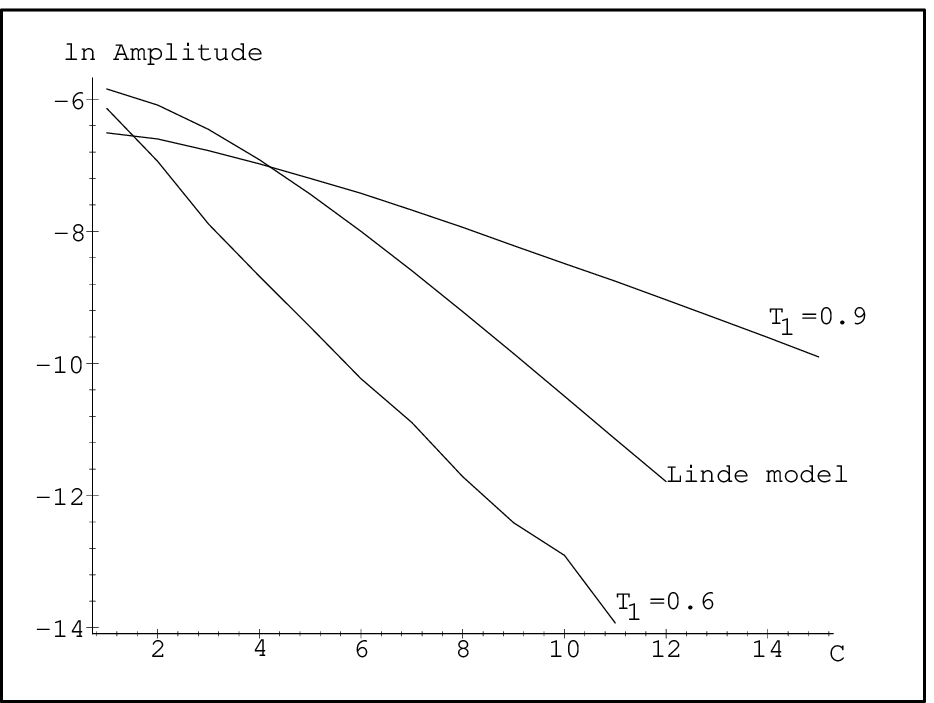}

\centerline{\em Fig 5: Damping of oscillation amplitude in comparison with
model of \cite{linde}}

\epsfbox[-80 0 500 210]{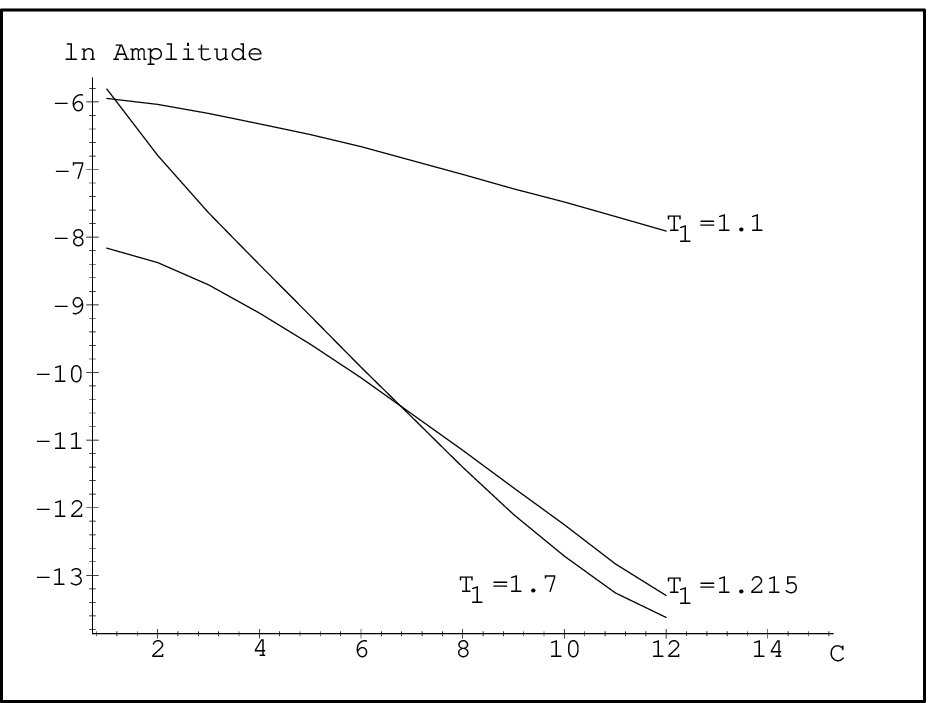}

\centerline{\em Fig 6: Damping of oscillation amplitude for additional starting
values}

\vspace{0.5cm}

These simulations consider $C$ to be independent of $T$. However, as we showed
in the last section, $C$ will usually depend on $T$. It is therefore necessary
to study the running of moduli with a non--constant $C$ and one can expect that
at least for $T$ being in the vicinity of the low energy  minimum, $C$ should
be large. 

The analysis confirms this: we choose $T_1=1.7$ as the high energy minimum
(starting value) and allowed $C$ to vary quadratically from 1 to 8 between this
initial and the final minimum like $C(T)=285 (T-1.7)^2+1$ and $C(T)=285
(T-1.23)^2+1$. Having $C=8$ at 
$T=1.7$ and $C=1$ at $T=1.23$ (the second case) results in negligible damping
of the  modulus oscillation. The final amplitude is even larger than for
constant $C=1$ although over the whole trajectory of $T$  the value of $C$ is
larger. 

The first choice, where $C$ becomes large as $T\rightarrow 1.23$, does show a
damping. It corresponds however not to the maximum value of $C$ along the
trajectory but to some average. The final amplitude is comparable to the case
of constant $C\approx 2.5$ for this example.

To achieve significant damping it is therefore a necessary condition to have a
large value of $C$ around the final minimum $T_0$, and a consistently large 
$C$ over the whole trajectory of the modulus $T$ would be sufficient.

\subsection{Explicit examples for a smooth transition}

As we have shown in the preceding sections, a model with sufficient modulus
damping necessarily will fulfill two criteria:~first, the high energy minimum
$T_1$ must not be close to a maximum of $V_0$ and second, $C^2(T_0)=C_0^2$ must
be large.

In addition it would simplify the analysis considerably if the trajectory of
the modulus would be along the real axis. In principle this is not an unnatural
requirement, since the theory obeys a symmetry $T\rightarrow \Tb$ and thus both
$V_0$ and $V_1$ have an extremum on the real axis in the imaginary
direction. This symmetry constitutes another example of how a symmetry of the
theory could correlate the minima of the low and high energy potentials. Again
the crucial point is to ensure that these extrema are indeed minima in the
imaginary direction. 

Unfortunately within the specified class of theories we did not succeed in
constructing such a `real' model, which has large
values of $C^2(T)$. However, models with
$T$ confined to the real axis and small
$C^2$ ($C^2 < 1$) can be constructed. One example is
$F(T)=M_\F/(\eta^6(T)j^{1/6}(T))$.

Relaxing the condition for $T$ to be real at all times we therefore attempt
to construct models with large $C^2$ and complex trajectory of $T$.
Furthermore we concentrate to on large final values $C_0^2$ (the necessary
condition), to get an
impression of the size of $C$ that can be expected.

Using the Yukawa--couplings \refs{yukawa} one gets for $C_0^2$ at the two
minima $T_0=0.81,\;1.23$ (for broken supersymmetry)
\bea
C^2(1.23)&=&\frac 1{q}\frac 1{1.23-T_1}\left ( -0.63\,n+4.49\,l+1.55 \right )\\
C^2(0.81)&=&\frac 1{q}\frac 1{0.81-T_1}\left ( -0.27\,n-6.78\,l+1.33 \right )\
. 
\eea

One can see, that these models are in principle able to explain large values of
$C^2$, which is due to the properties of the modular functions (especially
$j(T)$). The powers of the modular functions might however be
constrained by the fact that $V_1$ should have a minimum at finite $T$.
This is ensured if $F(T)\rightarrow \infty$ for $T_R\rightarrow \infty$. This
in turn gives the condition
\be
l>\frac q{12}n-\frac w{12}\label{limitl} \ .
\ee

One model which falls into this allowed region (modular weight $n=-4$) is then
\be
F(T)=\frac {M_\F}{\eta^{22}(T)j^{2/3}(T)}\ ,
\ee
with $C^2(0.81)\simeq 7/(0.81-T_1)/q$.

This is about as high as one can go if one stays within the orbifold limits on
the modular weight~\refs{n_bound} and \refs{limitl}.

\section{Summary and conclusion}

Our paper discusses the behaviour of string moduli during the preheating
period, the time after inflation when the energy density is dominated by the
coherent oscillations of $\F$.

We show that $\F$ can usually be integrated out to give an effective potential
for the modulus, with a crucial dependence on the Yukawa coupling.

For this paper we restrict ourselves to models which obey $T$--duality.  As
the low energy potential we use the one given by the standard gaugino
condensation picture. For these models it is found that a stabilization of
the modulus at high energy is possible for a limited range of the modular
weight $n$ of $\F$ ($n<3$) only. Within this limit, the reduction of the
initial modulus amplitude around 
the high energy minimum can be calculated and will be $10^5$ at
most. To be compatible with standard nucleosynthesis one would need a reduction
factor of about $10^{10}$.

Studying the further time evolution of the modulus, we consider the different
possibilities of how the modulus can settle to its low energy minimum.

First we construct explicit examples with coinciding low and high energy minima
at the self--dual 
points. In this case the initial stabilization will survive the subsequent
evolution. Nevertheless the reduction is not large enough to completely solve
the cosmological moduli problem, if the initial amplitude (at the end of
inflation) is of the order 1 (in Planck units). One therefore needs a
pre--stabilization by coinciding minima during inflation and preheating as
well.

Even then quantum fluctuations during
inflation will drive the modulus away from its minimum, so that the additional
reduction during preheating might be crucial to solve the moduli problem.

A stabilization at self--dual points implies unbroken supersymmetry in the
direction of the respective modulus. It can thus not work for all moduli, if
supersymmetry is broken by a modulus at all. Another mechanism to avoid modulus
oscillations, namely the smooth transition from high to low energy minimum
(induced by a mass term $C^2 H^2$ with large $C$) is studied next.

We found the modulus evolution in a more realistic (T--dual) potential to be
more complicated than in a quadratic potential.
If the high energy minimum is close to a low energy maximum the modulus
oscillation is not significantly damped, irrespectively of the value of
$C^2$. This might be quite relevant, since the usual T--dual potential for
broken supersymmetry has a saddle at the self--dual point $T=1$. 

The reduction
factor will also depend on the shape of  
the low energy potential along the trajectory of $T$. If $V$ is steep along
this trajectory, the reduction will be larger for the same values of $C$.

A nice feature of our framework is that it allows us to calculate $C$ 
from the model. Large values of $C$ are directly linked to steep couplings
(large gradient with respect to $T$).

Finally we analyze specific models and calculate the respective values of
$C(T)$. Our experience shows that it is difficult to construct examples which
fulfill all requirements, namely: a well defined high energy minimum, a large
value of $C^2(T)$ along the whole trajectory and (for convenience) a trajectory
along the real axis. Though modular functions like $j(T)$ are well suited to
produce steep couplings, we could not obtain values of $C^2$ larger than
$O(10)$. We believe that this in the upper region of what one can achieve. This
can be contrasted to the values of $C^2$ needed to solve the 
moduli problem, which have to be at least one or two orders of magnitude
larger. 

We believe that our work has serious implications for the cosmological moduli
problem. The additional stabilization of the modulus at its high energy
minimum during preheating will provide a large amplitude reduction which can
help in solving the moduli problem. As we have shown, 
a further damping of the oscillation can be achieved via the two possibilities
of coinciding minima and smooth transitions. For the second the
oscillation amplitude will be governed by the factor $C^2(T)$ and the shape of
the potential. However, it seems unlikely that the amplitude can be reduced
down to the nucleosynthesis bound. Nevertheless, given some additional small
reduction (e.g. by weak scale inflation), one might be able to achieve this. In
any case these results can significantly lessen the demands on other
mechanisms.

\vspace{0.5cm}

\noindent {\bf Acknowledgment} We thank Stephan Stieberger
for information about Yukawa--couplings in string theory. This work was
partially supported by the EC under contract no.~SC1-CT92-0789 and by the
Sonderforschungsbereich 375--95 `Research in Astroparticlephysics' of DFG. 

%%%%%%%%%%%%%%%%%%%%%%%%%%%%%%%%%%%%%%%%%%%%%%%%%%%%%%%%%%%%%%%%%
%%% Bibliography %%%%%%%%%%%%%%%%%%%%%%%%%%%%%%%%%%%%%%%%%%%%%%%%
%%%%%%%%%%%%%%%%%%%%%%%%%%%%%%%%%%%%%%%%%%%%%%%%%%%%%%%%%%%%%%%%%

%
\end{document}